# Quantitative Imaging of Single, Unstained Viruses with Coherent X-rays

Changyong Song,[1] Huaidong Jiang,[1] Adrian Mancuso,[1] Bagrat Amirbekian,[1] Li Peng,[2] Ren Sun,[2] Sanket S Shah,[3] Z. Hong Zhou,[4] Tetsuya Ishikawa,[5] and Jianwei Miao[1]

[1]Department of Physics and Astronomy, University of California, Los Angeles, CA 90095, USA.

[2]Department of Molecular and Medical Pharmacology, University of California, Los Angeles, CA 90095, USA.

[3]Department of Pathology and Laboratory Medicine, University of Texas-Houston Medical School, TX 77030, USA.

[4]Department of Microbiology, Immunology & Molecular Genetics, University of California, Los Angeles, CA 90095, USA.

[5]RIKEN SPring-8 Center, 1-1-1, Kouto, Sayo, Hyogo 679-5148, Japan.

**Since Perutz, Kendrew and colleagues unveiled the structure of hemoglobin and myoglobin based on X-ray diffraction analysis in the 1950s[1,2], X-ray crystallography has become the primary methodology used to determine the 3D structure of macromolecules. However, biological specimens such as cells, organelles, viruses and many important macromolecules are difficult or impossible to crystallize, and hence their structures are not accessible by crystallography. Here we report, for the first time, the recording and reconstruction of X-ray diffraction patterns from single, unstained viruses. The structure of the viral capsid inside a virion was visualized. This work opens the door for quantitative X-ray imaging of a broad range of specimens from protein machineries, viruses and organelles to whole cells. Moreover, our experiment is directly transferable to the use of X-ray free electron lasers[3], and**

**represents a major experimental milestone towards the X-ray imaging of single macromolecules[4-6].**

X-ray crystallography is one of the few techniques in history that has made a revolutionary impact upon such a broad range of fields including physics, chemistry, materials science, biology and medicine. However, the bottleneck of X-ray crystallography is the requirement of good quality crystals. Overcoming this major limitation requires the employment of different techniques. One promising approach currently under rapid development is X-ray diffraction microscopy (or coherent diffractive imaging) where the X-ray diffraction patterns of non-crystalline specimens are measured and then directly phased by the oversampling iterative algorithm[7-20]. However, due to the absence of the very large signal amplification that occurs in crystals, X-ray diffraction microscopy has so far been limited to the imaging of micron-sized or high-Z specimens[7-20]. Here we, for the first time, recorded and reconstructed X-ray diffraction patterns from single, unstained viruses that have molecular mass about three orders of magnitude smaller than previously investigated samples. By separating the diffraction pattern of the virus particles from that of their surroundings, we performed quantitative and high-contrast imaging of a single virion with a resolution of 22 nm. The structure of the viral capsid inside the virion was identified. With more brilliant synchrotron radiation sources[21] and future X-FELs[3], much higher resolutions should be achievable. Due to its quantitative capability, high image contrast and high spatial resolution, we anticipate that X-ray diffraction microscopy will become an important imaging technique for unveiling the structure of a broad range of biological systems including single protein machineries, viruses, organelles and whole cells.

The samples we studied are single murine herpesvirus-68 (MHV-68) virions. A herpesvirus virion has an asymmetric tegument and envelope outside of the icosahedrally symmetric capsid composed of defined numbers of subunits[22]. However, each virion may have a different size of tegument and envelope, and the viral capsid is not necessary in the center of the virion[23]. While cryo-electron microscopy can determine the capsid structure of herpesviruses by averaging over thousands of virus particles[24], the reconstructions of pleomorphic virions obtained by cryo-electron tomography are limited in low image contrast and high levels of noise[23]. The goal of this study is to perform quantitative and high-contrast imaging of single, unstained virions by using coherent X-rays. The MHV-68 virions were inactivated by UV light (500mJ) and chemically fixed by 3% glutaraldahyde. Unstained virions were suspended in methanol to regulate the concentration to about 20 virions/µL and supported on 30-nm thick silicon-nitride-membranes. Single, isolated virions were located with a high-resolution optical microscope and studied by the X-ray diffraction microscope.

The experiment was carried out on an undulator beamline at SPring-8. Figure 1 shows the schematic layout of the biological X-ray diffraction microscope. Unfocused, monochromatic X-rays with an energy of 5 keV were filtered by a 20 µm-diameter-pinhole, placed approximately 1 m upstream of the sample. A Si guard slit with beveled edges was positioned just in front of the sample to block the parasitic scattering from the upstream optical components. Coherent X-ray diffraction patterns were recorded by a liquid-nitrogen-cooled CCD camera with 1340×1300 pixels and a pixel size of 20 µm, located a distance of 1 m downstream of the sample. To obtain the diffraction pattern only from single virions, we measured two sets of diffraction intensities with the specimens in and out of the X-ray illumination, and then subtracted the two diffraction patterns. This procedure removed the unwanted scattering from specimens' surroundings and allowed us to perform quantitative and high-contrast imaging of single virions.

Figure 2a shows the diffraction pattern of a single, unstained virion which were added up from 3 independent diffraction patterns, each having a radiation dose of $3.5 \times 10^7$ Gy. Careful examination of the 3 diffraction patterns indicated that the radiation dose made minimum appreciable structure changes to the virion at this resolution. To significantly improve the signal-to-noise ratio (SNR) of the diffraction pattern, we integrated the diffraction intensities by binning 13×13 pixels into 1 pixel, and performed deconvolution to the integrated pattern[25]. The deconvolution process removed the effects of the finite pixel size of the CCD on the phase retrieval[25]. Figure 2a shows the characteristic ring structures of the diffraction pattern, reflecting the general round shape of the virion. Based on the diameter of the rings, the size of the virion was estimated to be ~ 200 nm. Intensity variation along the azimuthal angle is also visible in Fig. 2a, which is due to the internal structure of the virion.

The phase retrieval of the diffraction pattern was carried out by the guided hybrid-input-output algorithm (GHIO)[17,26] [Methods]. The GHIO started with 16 independent reconstructions of the diffraction pattern with different random phase sets as the initial input. Each reconstruction was iterated back and forth between real and reciprocal space while positivity and zero-density constraints were enforced. The iterative algorithm was guided towards minimizing the R-value, *i.e.* the difference between the measured and calculated Fourier modulus. The algorithm was terminated when the R-value could not be further improved and all the 16 independent reconstructions became very consistent.

Figure 2b shows the average of the 5 best images with the smallest R-values. To examine the reliability of the reconstruction, we performed another GHIO run of the X-ray diffraction pattern. Figures 3a and b show the final images from two independent GHIO runs. The difference between the two images was estimated to be 2.3% [Methods], indicating the robustness of the phase retrieval. We also took an SEM image of the same virion (Fig. 2c), on which a 5 nm thick Au film was deposited to remove the surface charging effect. While the SEM image only provides the surface morphology of the virion, the overall shape is in good agreement with the X-ray image. To characterize the internal structure, we took negative stain TEM images of similar virions, shown in Fig. 2d. The bright circular region in the TEM image represents the viral capsid which is tightly packed of the viral genome. The high-density region in the X-ray image (*i.e.* the darker area in Fig. 2b) shows features similar to the capsid structure of the TEM image. Compared with the thin-film-deposited SEM and negative stain TEM images, the X-ray diffraction image of a single, unstained virion shows the highest contrast as the background and surroundings of the virion were completely removed.

To quantify the X-ray image, we measured the incident and diffracted X-ray intensity, from which we calculated the absolute electron density of the virion [Methods]. Figure 4a shows the absolute electron density of the virion, where the yellow region represents the highest electron density with a size of ~100 nm. Figure 4b shows an AFM image of a similar virion, which was used to estimate the thickness profile of the virions. After taking the thickness into account, the electron density within the contour line was estimated to be approximately 1.3 times higher than the average density of the virion. It is reasonable to conclude that the high-density region represents the viral capsid of the single, unstained virion. To quantify the contrast of the viral capsid, we took a lineout across the capsid shown in Fig. 5. Due to the high-contrast ability of the X-ray diffraction microscope, the absolute electron density variations inside the capsid are visible (Fig. 5), which may be due to the packing of the viral genome.

Presently, the resolution of X-ray diffraction microscopy is limited by the coherent X-ray flux. With more brilliant synchrotron radiation sources[21], the resolution is ultimately limited by radiation damage to biological specimens[27,28]. While cryogenic cooling of biological specimens can alleviate the radiation damage problem[29], recent studies have shown that the highest resolution attainable for imaging the 3D pleomorphic structure of biological specimens is ~5 nm[27,28]. For the imaging of large protein molecules having identical copies, the resolution may be further improved by employing extremely intense and ultrafast X-ray pulses such as X-FELs[3]. Computer modeling and experimental results have both indicated that significant damage occurs only after an ultrafast X-ray pulse ($\leq 25$ fs) traverses a specimen[4,6]. By using computer simulations, it has been shown that approximately $10^5 - 10^6$ identical copies of single large protein (*i.e.* molecular mass > 100 kDa) can lead to the 3D structure of the molecules at the near atomic resolution[5].

In summary, coherent X-ray diffraction patterns were obtained from a single, unstained herpesvirus virion, and then directly inverted to yield quantitative and high-contrast electron density maps with a resolution of 22 nm. The quantitative structure of the viral capsid inside the virion was visualized. While the present resolution is limited by the coherent X-ray flux, higher resolutions should be achievable by using more brilliant synchrotron radiation sources[21]. This work hence opens the door for broad application of X-ray diffraction microscopy to biological specimens ranging from single large protein machineries, viruses and cellular organelles to whole cells. As tabletop soft X-ray diffraction microscopy based on the high harmonic generation and soft X-ray laser sources has recently been demonstrated[19], the combination of this work with compact X-ray sources will make this imaging technique more accessible to the biology community. Finally, X-FELs are undergoing rapid development worldwide[3]. The major driving force for these large-scaled coherent X-ray sources is the potential of imaging single biomolecules. The first recording and reconstruction of X-ray diffraction patterns from single, unstained virions hence represents a major experimental milestone towards the ultimate goal of imaging large protein complexes[4,5].

**Methods**

**Virion purification** Baby Hamster Kidney (BHK-21) cells were infected with MHV-68 at a multiplicity of infection (MOI) of 0.01. When the cultured cells exhibited 90% cytopathic effect, they were pelleted by centrifugation at 1,000 x g for 10 min at 4°C, and the extracellular media containing MHV-68 virions was retained for further purification. Briefly, the supernatant was centrifuged at 23,000 x g for 2 hours to pellet virions. The resultant pellet was re-suspended and overlaid on a 15–35% potassium tartrate and 30–0% glycerol gradient, and centrifuged at 55,000 x g for I hr. The visible bands were collected and checked for virions using negative stain transmission electron microscopy. The virion-containing band was re-suspended in phosphate buffered saline (PBS) and concentrated by centrifugation at 60,000 x g for 1 hr.

**Phase Retrieval** The GHIO algorithm began with 16 independent reconstructions of each diffraction pattern with random initial phase sets as the initial input. Each reconstruction was iterated between real and reciprocal space via forward and inverse fast Fourier transformations. In real space, the sample density outside a support and the negative real or imaginary part of the density inside the support were slowly pushed close to zero. The support, a region within which the sample image was confined, was a rectangular shape with its size estimated from the linear oversampling ratio[30]. In reciprocal space, the Fourier modulus remains unchanged while the phase of each pixel was updated with each iteration. After 3000 iterations, 16 images were reconstructed, which was defined as the 0$^{th}$ generation. An R-value was calculated for each image, defined as,

$$R = \frac{\sum_{k_x,k_y} \left\| F(k_x,k_y) \right| - \left| G(k_x,k_y) \right\|}{\sum_{k_x,k_y} \left| F(k_x,k_y) \right|} \quad (1)$$

where $\left| F(k_x,k_y) \right|$ and $\left| G(k_x,k_y) \right|$ are the measured and calculated Fourier modulus, $k_x$ and $k_y$ are the coordinates in reciprocal space. A seed image was selected for the reconstruction with the smallest R-value. By multiplying the seed with each of the 16 images and taking the square root of the product, a new set of 16 images was obtained, which was used as the initial inputs for the next generation. We repeated the procedure for the next generation, and after the 6$^{th}$ generation, the 16 reconstructed images became very consistent. The final reconstruction was chosen from the average of best 5 images with the smallest R-values.

Figures 3a and b show the final reconstructed images from the two independent GHIO runs. The reconstruction error ($R_{err}$) was calculated by

$$R_{err} = \frac{\sum_{x,y} \left| \rho_1(x,y) - \rho_2(x,y) \right|}{\sum_{x,y} \left| \rho_1(x,y) + \rho_2(x,y) \right|} \quad (2)$$

where $\rho_1(x,y)$ and $\rho_2(x,y)$ represent the two final reconstructed images. $R_{err}$ between Figs. 3a and b was estimated to be ~ 2.3%.

**Calculation of the Absolute Electron Density** The absolute electron density of the virion was determined by,

$$I(0,0) = I_0 \, r_e^2 \left| F(0,0) \right|^2 \frac{\Delta s}{r^2} \quad (3)$$

where $I(0,0)$ represents the number of diffracted X-ray photons in the forward direction (*i.e.* within the centro-pixel), $I_0$ is the incident X-ray flux per unit area, $r_e$ is the classical electron radius, $|F(0,0)|$ represents the total number of electrons in the virion, $\Delta s$ is the area of the centro-pixel, and $r$ is the distance from the sample to the CCD. Although $I(0,0)$ cannot be experimentally measured due to a beamstop, our recent results have shown that as long as the missing data is confined within the centro-speckle, the missing data can be reliably recovered from the measured diffraction intensities alone[31].

**ACKNOWLEDGEMENTS**

We thank Y. Nishino and Y. Kohmura for data acquisition. This work was supported by the U.S. Department of Energy, Office of Basic Energy Sciences under the contract number, DE-FG02-06ER46276 and the U.S. National Science Foundation, Division of Materials Research (DMR-0520894). Use of the RIKEN beamline (BL29XUL) at SPring-8 was supported by RIKEN.

**AUTHOR CONTRIBUTIONS**

J.M., C.J., H.J and A.M. designed research; R.S., Z.H.Z., L.P. and S.S.S. prepared virion specimens and took TEM images of the virions; C.S., H.J. A.M., T.I. and J.M. performed research; C.S. H.J., A.M., B.A. and J.M. analyzed the data; C.S., H.J., A.M., S.S.S., Z.H.Z. and J.M. wrote the manuscript.

**COMPETING INTERESTS STATEMENT**

The authors declare no competing financial interests. Correspondence and requests for materials should be addressed to J. M. (miao@physics.ucla.edu).

**Figure Captions**

**Figure 1** Schematic layout of the X-ray diffraction microscope. A 20-um-pinhole was used to define the incident X-ray wave. The virion specimen was positioned at a distance of 1 m

from the pinhole. A silicon guard slit with beveled edges was used to eliminate the parasitic scattering from the pinhole. The oversampled diffraction pattern, recorded on a liquid-nitrogen-cooled CCD camera, was directly inverted to a high-contrast image using an iterative algorithm.

**Figure 2** X-ray diffractive imaging of single herpesvirus virions. (**a**) X-ray diffraction pattern obtained from a single, unstained virion. (**b**) High-contrast image reconstructed from (**a**) where the background and the surroundings of the virion were completely removed. (**c**) SEM image of the same virion. (**d**) Negative stain TEM image of a similar herpesvirus virion.

**Figure 3** (**a**), (**b**) The final reconstructed images from two independent GHIO runs. Each image was obtained by averaging the 5 best images with the smallest R-values.

**Figure 4** (**a**) Quantitative characterization of the reconstructed electron density map of the herpesvirus virion. The dashed line was used to obtain the lineout shown in Fig. 5. (**b**) AFM image of a similar virion, showing the thickness profile of a similar virion.

**Figure 5** Lineout across the viral capsid (*i.e.* along the dashed line in Fig. 4a), showing the distribution of the absolute electron density. The density variations inside the capsid may be due to the packing of the viral genome.

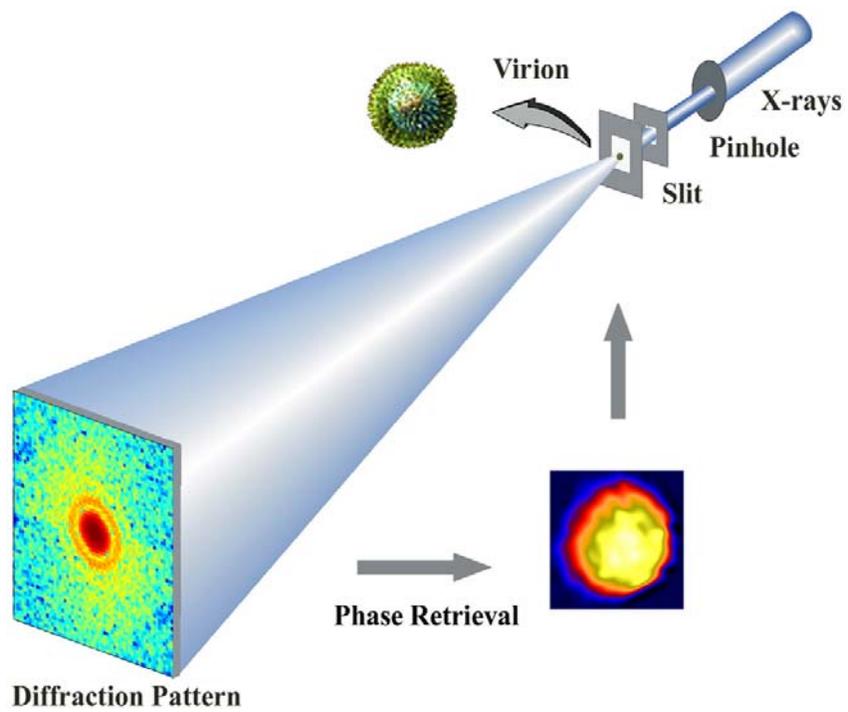

FIG. 1

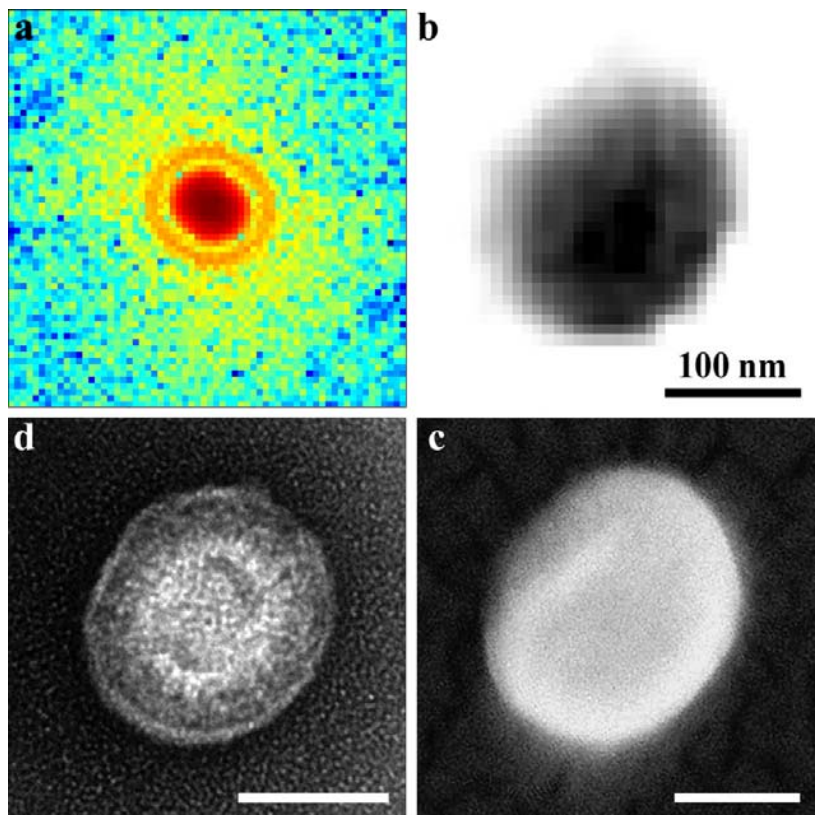

FIG. 2

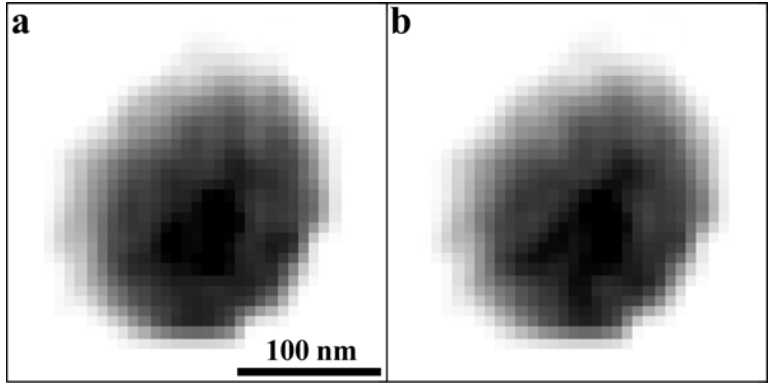

FIG. 3

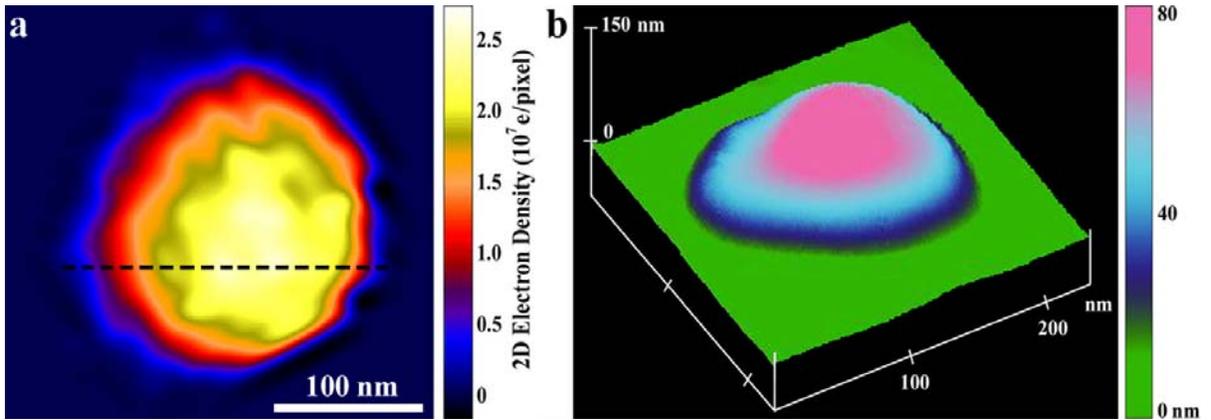

FIG. 4

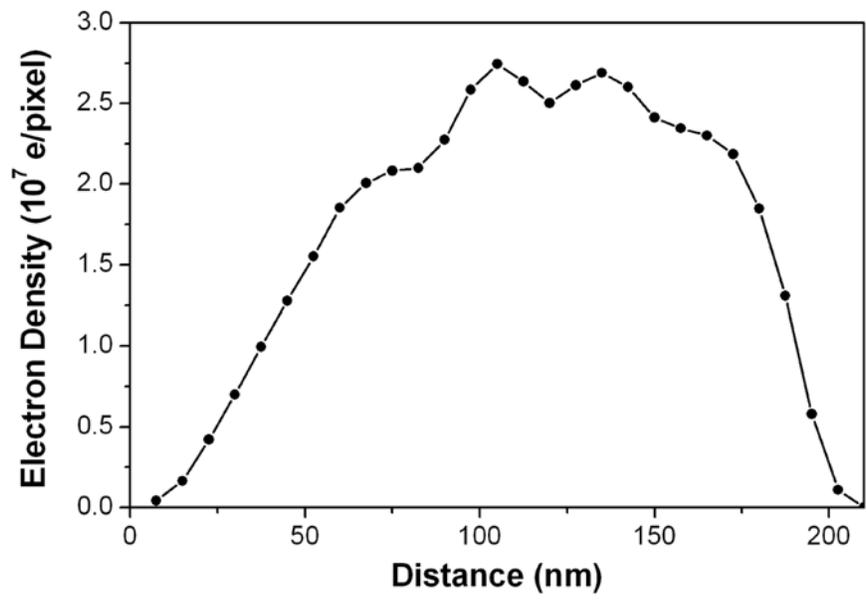

FIG. 5